\documentclass[onecolumn,showpacs,11pt]{revtex4}

\usepackage{graphicx}
\usepackage{dcolumn}
\usepackage{bm}

\input epsf

\begin{document}

\title{\Large Accretion of Dark Energy onto Higher Dimensional Charged BTZ Black Hole}

\author{\bf Ujjal Debnath\footnote{ujjaldebnath@gmail.com ,
ujjal@iucaa.ernet.in}}

\affiliation{Department of Mathematics, Indian Institute of
Engineering Science and Technology, Shibpur, Howrah-711 103,
India.\\}

\date{\today}

\begin{abstract}
In this work, we have studied the accretion of $(n+2)$-dimensional
charged BTZ black hole (BH). The critical point and square speed
of sound have been obtained. The mass of the BTZ BH has been
calculated and we have observed that the mass of the BTZ BH is
related with square root of the energy density of dark energy
which accretes onto BH in our accelerating FRW universe. We have
assumed modified Chaplygin gas (MCG) as a candidate of dark energy
which accretes onto BH and we have found the expression of BTZ BH
mass. Since in our solution of MCG, this model generates only
quintessence dark energy (not phantom) and so BTZ BH mass
increases during the whole evolution of the accelerating universe.
Next we have assumed 5 kinds of parametrizations of well known
dark energy models. These models generate both quintessence and
phantom scenarios i.e., phantom crossing models. So if these dark
energies accrete onto the BTZ BH, then in quintessence stage, BH
mass increases upto a certain value (finite value) and then
decreases to a certain finite value for phantom stage during whole
evolution of the universe. We have shown these results
graphically.
\end{abstract}


\maketitle

\section{Introduction}

In recent years, the type Ia Supernovae and Cosmic Microwave
Background (CMB) \cite{Perlmutter,Riess} observations suggest that
our universe is currently in the phase of accelerated expansion.
This acceleration is caused by some unknown matter which has the
property that positive energy density and negative pressure
satisfying $\rho+3p<0$ is known as ``dark energy'' (DE)
\cite{Briddle,Spergel,Peebles,Cald}. The simplest candidate of
dark energy is the cosmological constant which is characterized by
the equation of state $p=w\rho$ with $w=-1$. Many other
theoretical models have been proposed to explain the accelerated
expansion of the universe. Another candidate of dark energy is
quintessence satisfying $-1<w<-1/3$ \cite{Peebles,Cald}. When
$w<-1$, it is known as phantom energy. Distinct data on supernovas
showed that the presence of phantom energy with $-1.2 < w < -1$ in
the Universe is highly likely \cite{Alam2}. Several models for the
explanation of dark energy were suggested. These usually include
k-essence, dilaton, DBI-essence, Hessence, tachyon, Chaplygin gas,
etc \cite{Arme, gas, Gum, Mart, Wei, Sen, Cald1, Kamen, Cop}.\\

In Newtonian theory, the problem of accretion of matter onto the
compact object was formulated by Bondi \cite{Bondi}. The equations
of motion for steady-state spherical symmetric flow of matter into
or out of a condensed object (e.g. neutron stars, `black holes',
etc.) are discussed by Michel \cite{Michel} and also obtained
analytic relativistic accretion solution onto the static
Schwarzschild black hole. The accretion of phantom energy onto a
static Schwarzschild black hole was first proposed by Babichev et
al \cite{Babichev,Babichev1} and established that black hole mass
will gradually decrease due to strong negative pressure of phantom
energy and finally all the masses tend to zero near the big rip
where it will disappear. Jamil \cite{Jamil} has investigated the
accretion of phantom like variable modified Chaplygin gas onto
Schwarzschild black hole and also showed that mass of the black
hole will decreases for dark energy accretion and otherwise will
increases. Also the accretion of dark energy onto the more general
Kerr-Newman black hole was studied by Madrid et al \cite{Pedro}
and Bhadra et al \cite{Bhadra}. Till now, several authors
\cite{Chak,Maj,Nayak,Dwiv,Lima,Sharif,Sharif1,Sun,Kim,Mar,Sharif2,Rod,Abhas,Abhas1,Mar1}
have discussed the accretion of several candidates of dark energy
onto black holes.\\

Recently, there has been a growing interest to study the black
hole (BH) solution in (2+1)-dimensions. These BH solutions have
all the typical properties that can be found in (3+1) or higher
dimensions, such as horizons, Hawking temperature and
thermodynamics. The discovery and investigation of the
(2+1)-dimensional BTZ (Banados-Teitelboim-Zanelli) black holes
\cite{Bana,Bana1,Empar} organizes one of the great advances in
gravity. Jamil and Akbar \cite{Jamil1} have investigated the
thermodynamics of phantom energy accreting onto a BTZ BH. Abhas
\cite{Abha} investigated the phantom energy accretion onto 3D
black hole formulated in Einstein-Power-Maxwell theory. The
accretion of phantom energy onto Einstein-Maxwell-Gauss-Bonnet
black holes was studied in \cite{J}. They showed that the
evolution of the black hole mass was independent of its mass and
depends only on the energy density and pressure of the phantom
energy. Interest in the BTZ black hole has recently heightened
with the discovery that the thermodynamics of higher dimensional
black holes \cite{Kim1,Has}.  Also, non-static charged BTZ like
black holes in $(n+1)$-dimensions have been considered by Ghosh et
al \cite{Ghosh}, which in the static limit, for $n = 2$, reduces
to (2+1)-dimensional BTZ black hole solutions. John et al
\cite{John} examined the steady-state spherically symmetric
accretion of relativistic fluids, with a polytropic equation of
state, onto a higher dimensional Schwarzschild black hole. Also
charged BTZ-like black holes in higher dimensions have been
studied by Hendi \cite{Hendi}. There are also charged, rotating,
regular extensions of the BTZ black hole solutions
\cite{Shu,Bir,Chakra,Rubio,Lar,Seta,Ak,Ak1} available in the
literature by employing nonlinear Born-Infeld
electrodynamics to eliminate the inner singularity \cite{Maz}.\\

In section II, we assume the $(n+2)$-dimensional charged BTZ black
hole (BH) in presence of dark energy filled universe. The critical
point has been obtained. If dark energy accretes onto the BTZ BH,
the rate of change of mass of the black hole is expressed in terms
of the density and pressure of dark energy and also find the
expression of BH mass in terms of density. In our previous work,
we have investigated accretions of various types of dark energies
(including some kinds of parametrizations of dark energy) onto
Morris-Thorne wormhole \cite{Debn}. Our main motivation of the
work is to examine the natures of the mass of the black hole
during accelerating expansion of the FRW universe if several kinds
of dark energies accrete around the BH. In section III, we have
assumed some versions of dark energy like modified Chaplygin gas
(MCG) and some kinds of parametrizations of dark energy
candidates. The mass of the BTZ BH has been calculated for all
types of dark energies and its natures have been analyzed during
evolution of the universe. Finally, we give some concluding
remarks of the whole work in section IV.\\

\section{Accretion Phenomena of Higher Dimensional Charged BTZ Black Hole}

In recent years there has been increasing interest about black
hole solutions whose matter source is power Maxwell invariant,
i.e., $(F_{\mu\nu}F^{\mu\nu})^s$ \cite{Has,Hendi}, where $s$ is
the power of non-linearity. In the special case $(s = 1)$, it can
reduces to linear electromagnetic field. In addition, in
$(n+2)$-dimensional gravity, for the special choice $s = (n+2)/4$,
matter source yields a traceless Maxwell's energy-momentum tensor
which leads to conformal invariance, which is the analogues of the
four dimensional Reissner-Nordstrom solutions in higher dimensions
\cite{Hendi,Has1}. Also, it is valuable to find and analyze the
effects of exponent $s$ on the behavior of the new solutions, when
$s = (n+1)/2$. In this case the solutions are completely different
from another cases $(s \ne (n+1)/2)$.\\

The $(n+2)$-dimensional action in which gravity is coupled to
nonlinear electrodynamics field is given by \cite{Hendi}

\begin{equation}
S=\frac{1}{16\pi} \int d^{n+2}x \sqrt{-g}\left[R+2\Lambda-(\alpha
F)^{s} +L_{m} \right]
\end{equation}

where $R$ is scalar curvature, $\Lambda$ refers to the positive
cosmological constant which is in general equal to
$\frac{n(n+1)}{2l^{2}}$ for asymptotically AdS solutions, in which
$l$ is a scale length factor, $\alpha$ is a constant and
$s=(n+1)/2$ gives BTZ-like solutions. Varying the action (1) with
respect to the metric $g_{\mu\nu}$ and the gauge field $A_{\mu}$,
(with $s = (n+1)/2$) the field equations are obtained as

\begin{equation}
G_{\mu\nu}-\Lambda g_{\mu\nu}=T_{\mu\nu}^{(m)}+T_{\mu\nu}^{(EM)}
\end{equation}
Here,
\begin{equation}
T_{\mu\nu}^{(m)}=(\rho+p)u_{\mu}u_{\nu}+pg_{\mu\nu}
\end{equation}
is the energy-momentum tensor for matter. Here $\rho$ and $p$ are
the energy density and pressure of the matter while
$u^{\mu}=(u^{0},u^{1},0,0,...,0)$ is the velocity vector of the
fluid flow satisfying $u_{\mu}u^{\mu}=-1$. Also $u^{1} = u$ is the
radial velocity of the flow. Also
\begin{equation}
T_{\mu\nu}^{(EM)}=\alpha(\alpha F)^{\frac{n-1}{2}}
\left(\frac{1}{2}~g_{\mu\nu}F-nF_{\mu\lambda}F_{\nu}^{\lambda}
\right)
\end{equation}
is the energy-momentum tensor for electro-magnetic field and
\begin{equation}
\partial_{\mu}\left(\sqrt{-g} ~F^{\mu\nu}(\alpha F)^{\frac{n-1}{2}}\right)=0
\end{equation}

Let us consider static spherically symmetric $(n+2)$-dimensional
charged BTZ black hole metric given by \cite{Hendi}
\begin{eqnarray}
ds^2=-f(r)dt^{2}+\frac{1}{f(r)}~dr^{2}+r^{2}\sum_{i=1}^{n}d\phi_{i}^{2}
\end{eqnarray}
Here, $f(r)$ is termed as the lapse function, which is obtained as
\cite{Hendi}
\begin{equation}
f(r)=\frac{r^{2}}{l^{2}}-r^{1-n}\left[M+2^{\frac{n+1}{2}}Q^{n+1}\ln\left(\frac{r}{l}\right)
\right]
\end{equation}
where $M$ is the mass and $Q$ is the charge of the BTZ black hole.
Here $\sqrt{-g}=r^{n}$. Using $u_{\mu}u^{\mu}=-1$, we get
$g_{00}u^{0}u^{0}+g_{11}u^{1}u^{1}=-1$ (since $u^{0}$ and $u^{1}$
are the non-zero components of velocity vector), so we can obtain
$(u^{0})^{2}=\frac{(u^{1})^{2}+f(r)}{f^{2}(r)}$ and since $u^{1}=u$,
so we have $u_{0}=g_{00}u^{0}=\sqrt{u^{2}+f(r)}$.\\

A proper dark-energy accretion model for BTZ black hole should be
obtained by generalizing the Michel's theory \cite{Michel}. Such a
generalization has been already performed by Babichev et al
\cite{Babichev,Babichev1} for the case of dark-energy accretion
onto Schwarzschild black holes. We shall follow now the procedure
used by Babichev et al \cite{Babichev,Babichev1}. We assume that
the in-falling dark energy fluid does not disturb the spherical
symmetry of the black hole. The relativistic Bernoulli's equation
after the time component of the energy-momentum conservation law
$T^{\mu\nu}_{;\nu}=0$, we obtain (consider steady state condition
and spherically symmetric)

\begin{eqnarray}
\frac{d}{dr}~(T_{0}^{~1}\sqrt{-g})=0
\end{eqnarray}
which provides the first integral,
\begin{eqnarray}
(\rho+p)u_{0}u^{1}\sqrt{-g}=C_{1}
\end{eqnarray}
i.e.,
\begin{eqnarray}
ur^{n}(\rho+p)\sqrt{u^{2}+f(r)}=C_{1}
\end{eqnarray}

where the integration constant $C_{1}$ has the dimension
of the energy density.\\

Moreover, the second integration of motion is obtained from the
projection of the conservation law for energy-momentum tensor onto
the fluid four-velocity, $u_\mu T^{\mu \nu}_{;\nu}=0$, which gives
\begin{eqnarray}
u^{\mu}\rho_{,\mu}+(\rho+p)u^{\mu}_{;\mu}=0
\end{eqnarray}
which yields
\begin{eqnarray}
ur^{n}~exp\left[\int_{\rho_{\infty}}^{\rho_{h}}\frac{d\rho}{\rho+p}
\right]=-A
\end{eqnarray}
where $A$ is integration constant and the associated minus sign is
taken for convenience. Also $\rho_{h}$ and $\rho_{\infty}$ are the
energy densities at the BTZ horizon and at infinity respectively.
Combining these two, we obtain,

\begin{eqnarray}
(\rho+p)\sqrt{u^{2}+f(r)}~exp\left[-\int_{\rho_{\infty}}^{\rho_{h}}\frac{d\rho}{\rho+p}
\right]=C_{2}
\end{eqnarray}

where, $C_{2}=-C_{1}=\rho_{\infty}+p(\rho_{\infty})$. Further the
value of the constant $C_{2}$ can be evaluated for different dark
energy models. \\

The equation of mass flux $J^{\mu}_{;\mu}=0$ is given by
$\frac{d}{dr}~(J^{1}\sqrt{-g})=0$, which integrates to $\rho
u^{1}\sqrt{-g}=A_{1}$ yields
\begin{eqnarray}
\rho ur^{n}=A_{1}
\end{eqnarray}
where, $A_{1}$ is the integration constant. From (10) and (14), we
obtain,
\begin{eqnarray}
\frac{\rho+p}{\rho}~\sqrt{u^{2}+f(r)}=\frac{C_{1}}{A_{1}}=C_{3}
\end{eqnarray}
Let,
\begin{eqnarray}
V^{2}=\frac{d\ln(\rho+p)}{d\ln\rho}-1
\end{eqnarray}
So from (14) and (15), we obtain
\begin{eqnarray}
\left[V^{2}-\frac{u^{2}}{u^{2}+f(r)}
\right]\frac{du}{u}-\left[nV^{2}-\frac{rf'(r)}{2(u^{2}+f(r))}
\right]\frac{dr}{r}=0
\end{eqnarray}

It is evident that if one or the other of the bracketed factors in
(17) vanishes one has a turn-around point, and the solutions are
double-valued in either $r$ or $u$. Only solutions that pass
through a critical point correspond to material falling into (or
flowing out of) the object with monotonically increasing velocity
along the particle trajectory. The critical point of accretion is
located at $r=r_{c}$ which is obtained by taking the both
bracketed factors in Eq. (17) to be zero. So at the critical
point, we have

\begin{eqnarray}
V_{c}^{2}=\frac{u_{c}^{2}}{u_{c}^{2}+f(r_{c})}
\end{eqnarray}
and
\begin{eqnarray}
nV_{c}^{2}=\frac{r_{c}f'(r_{c})}{2(u_{c}^{2}+f(r_{c}))}
\end{eqnarray}
Here, subscript $c$ refers to the critical quantity and $u_{c}$ is
the critical speed of flow at the critical points. From above two
expressions, we have
\begin{eqnarray}
u_{c}^{2}=\frac{r_{c}}{2n}~f'(r_{c})
\end{eqnarray}
At the critical point, the sound speed can be determined by
\begin{eqnarray}
c_{s}^{2}=\left. \frac{d
p}{d\rho}\right|_{r=r_{c}}=\frac{C_{3}V_{c}(V_{c}^{2}+1)}{u_{c}}
-1
\end{eqnarray}
We mentioned that the physically acceptable solutions of the above
equations are obtained if $u_{c}^{2}>0$ and $V_{c}^{2}>0$ which
leads to
\begin{eqnarray}
u_{c}^{2}>-f(r_{c})~~\text{and}~~f'(r_{c})>0
\end{eqnarray}
i.e.,
\begin{eqnarray}
u_{c}^{2}>-\frac{r_{c}^{2}}{l^{2}}+r_{c}^{1-n}\left[M+2^{\frac{n+1}{2}}Q^{n+1}\ln\left(\frac{r_{c}}{l}\right)
\right]
\end{eqnarray}
and
\begin{eqnarray}
2r_{c}^{n+1}+l^{2}\left[(n-1)M+ 2^{\frac{n+1}{2}}Q^{n+1}\left\{
(n-1)\ln\left(\frac{r_{c}}{l}\right)-1 \right\} \right]>0
\end{eqnarray}

For linear equation of state $p=w\rho$, we obtain $c_{s}^{2}=w$
and $V_{c}^{2}=0$ and from (18), we obtain $u_{c}=0$. From (20),
we see that the critical point occurs at the point $r_{c}$ where
$r_{c}$ can be found from the equation
\begin{equation}
r_{c}^{n+1}=l^{2}\left[M+2^{\frac{n+1}{2}}Q^{n+1}\ln\left(\frac{r_{c}}{l}\right)
\right]
\end{equation}

The rate of change of mass $\dot{M}$ of the BTZ black hole is
computed by integrating the flux of the dark energy over the
$n$-dimensional volume of the black hole and given by \cite{John}
\begin{eqnarray}
\dot{M}=-\frac{2\pi^{\frac{n+1}{2}}}{\Gamma(\frac{n+1}{2})}~r^{n}T_{0}^{1}
\end{eqnarray}
Using equations (12) and (13), the above equation can be written
as
\begin{eqnarray}
\dot{M}=\frac{2\pi^{\frac{n+1}{2}}}{\Gamma(\frac{n+1}{2})}~A(\rho_{\infty}+p(\rho_{\infty}))
\end{eqnarray}
If we neglect the cosmological evolution of $\rho_{\infty}$ then
from (26) we obtain the mass of the black hole as
\begin{eqnarray}
M=M_{0}+\frac{2\pi^{\frac{n+1}{2}}}{\Gamma(\frac{n+1}{2})}~A(\rho_{\infty}+p(\rho_{\infty}))(t-t_{0})
\end{eqnarray}
where $M_{0}$ is the initial mass corresponding to the initial
time $t_{0}$. The result (26) is also valid for any equation of
state $p=p(\rho)$, thus we can write
\begin{eqnarray}
\dot{M}=\frac{2\pi^{\frac{n+1}{2}}}{\Gamma(\frac{n+1}{2})}~A(\rho+p)
\end{eqnarray}

We see that the rate for the BTZ black hole exotic mass due to
accretion of dark energy becomes exactly the positive to the
similar rate in the case of a Schwarzschild black hole,
asymptotically. Since the BTZ black hole is static, so the mass of
the black hole depends on $r$ only. When some fluid accretes
outside black hole, the mass function $M$ of the black hole is
considered as a dynamical mass function and hence it should be a
function of time also. So $\dot{M}$ is time dependent and the
increasing or decreasing of the black hole mass $M$ sensitively
depends on the nature of the fluid which accretes upon the black
hole. If $\rho+p<0$ i.e., for phantom dark energy accretion, the
mass of the black hole decreases but if $\rho+p>0$ i.e., for
quintessence dark energy accretion, the mass of the black hole
increases.

\section{Dark Energy Accretes upon BTZ black hole}

In the following, we shall assume different types of dark energy
models such as modified Chaplygin gas and some parameterizations
of dark energy models. The natures of mass function of black hole
will be analyzed for present and future stages of expansion of the
universe when the dark energies are accreting upon BTZ black hole.

\subsection{Modified Chaplygin Gas}

We consider the background spacetime is spatially flat represented
by the homogeneous and isotropic FRW model of the universe which
is given by
\begin{equation}
ds^{2}=-dt^{2}+a^{2}(t)\left[dr^{2}+r^{2}(d\theta^{2}+sin^{2}\theta
d\phi^{2}) \right]
\end{equation}

where $a(t)$ is the scale factor. The Einstein's equations for FRW
universe are (choosing $8\pi G=c=1$)
\begin{eqnarray}
H^2 = \frac{1}{3} \rho~,
\end{eqnarray}
\begin{eqnarray}
\dot{H}=-\frac{1}{2}\left(p + \rho \right)
\end{eqnarray}

Conservation equation is given by
\begin{eqnarray}
\dot{\rho}+3H(\rho+p)=0
\end{eqnarray}

where $H=\frac{\dot{a}}{a}$ is the Hubble parameter. Now assume
the modified Chaplygin gas (MCG) \cite{Deb} as dark energy model,
whose EoS is $p=w\rho-\frac{B}{\rho^{\alpha}}$ ,
($B>0,~0\le\alpha<1$). For MCG model, we obtain the solution of
$\rho$ as
\begin{equation}
\rho=\left[\frac{B}{1+w}+\frac{C}{a^{3(1+w)(1+\alpha)}}
\right]^{\frac{1}{1+\alpha}}
\end{equation}
where $C>0$ is an arbitrary integration constant. From above, we
can obtain the present value of the energy density
$\rho_{0}=\left[\frac{B}{1+w}+C \right]^{\frac{1}{1+\alpha}}$. For
MCG model, we obtain $c_{s}^{2}=w+\frac{\alpha
B}{\rho^{\alpha+1}}$ and
$V_{c}^{2}=\frac{(\alpha+1)B}{(1+w)\rho^{\alpha+1}-B}$. Using
equations (29), (31) and (33), we have
\begin{eqnarray}
\dot{M}=-\frac{2\pi^{\frac{n+1}{2}}A}{\sqrt{3}~\Gamma(\frac{n+1}{2})}~\frac{\dot{\rho}}{\sqrt{\rho}}
\end{eqnarray}
which integrates to yield
\begin{eqnarray}
M=M_{0}-\frac{4\pi^{\frac{n+1}{2}}A}{\sqrt{3}~\Gamma(\frac{n+1}{2})}~(\sqrt{\rho}-\sqrt{\rho_{0}})
\end{eqnarray}
where, $M_{0}$ is the present values of the BTZ black hole mass.
In the late stage of the universe i.e., $a$ is very large
$(z\rightarrow -1)$, the mass of the black hole will be
\begin{eqnarray}
M=M_{0}+\frac{2\pi^{\frac{n+1}{2}}A}{\sqrt{3}~\Gamma(\frac{n+1}{2})}~\sqrt{\rho_{0}}
\end{eqnarray}

If we put the solution $\rho$ from equation (34) in equation (36),
the mass of black hole $M$ can be expressed in terms of scale
factor $a$ and then use the formula of redshift $z=\frac{1}{a}-1$,
$M$ will be in terms of redshift $z$, i.e.,

\begin{eqnarray}
M=M_{0}-\frac{4\pi^{\frac{n+1}{2}}A}{\sqrt{3}~\Gamma(\frac{n+1}{2})}
~\left\{\left[\frac{B}{1+w}+C(1+z)^{3(1+w)(1+\alpha)}
\right]^{\frac{1}{2(1+\alpha)}} - \left[\frac{B}{1+w}+C
\right]^{\frac{1}{2(1+\alpha)}} \right\}
\end{eqnarray}

Now $M$ vs $z$ is drawn in figure 1. Since our solution for MCG
model generates only quintessence, so from the figure, we see that
the mass $M$ of the BTZ BH always increases with $z$ decreases. So
we conclude that the mass of the BTZ BH increases if the MCG
accretes onto the BTZ BH.

\begin{figure}
\includegraphics[height=2.0in]{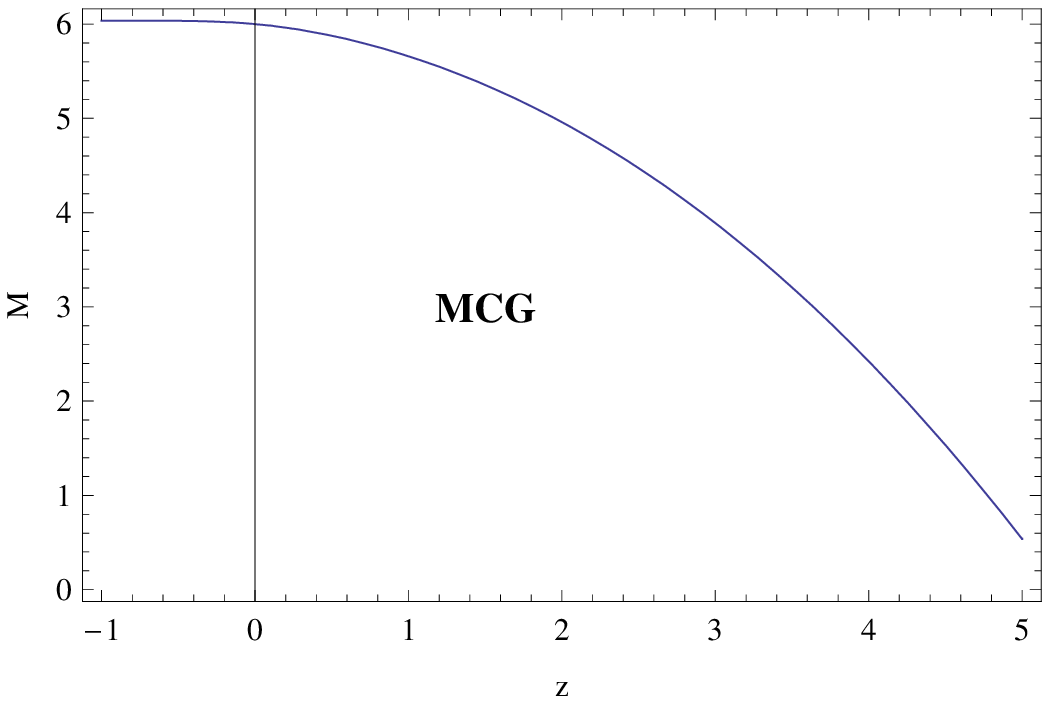}~~~~
\includegraphics[height=2.0in]{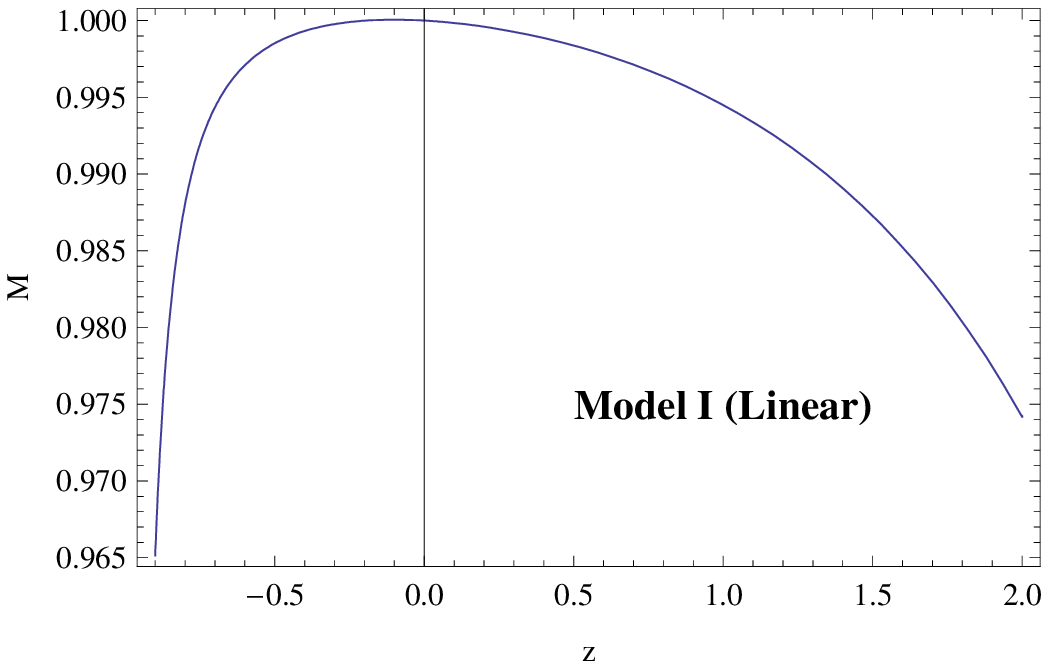}
\vspace{4mm}
~~~~~Fig.1~~~~~~~~~~~~~~~~~~~~~~~~~~~~~~~~~~~~~~~~~~~~~~~~~~~~~~~~~~~~~~~~~~~~~~Fig.2\\
\vspace{4mm}
\includegraphics[height=2.0in]{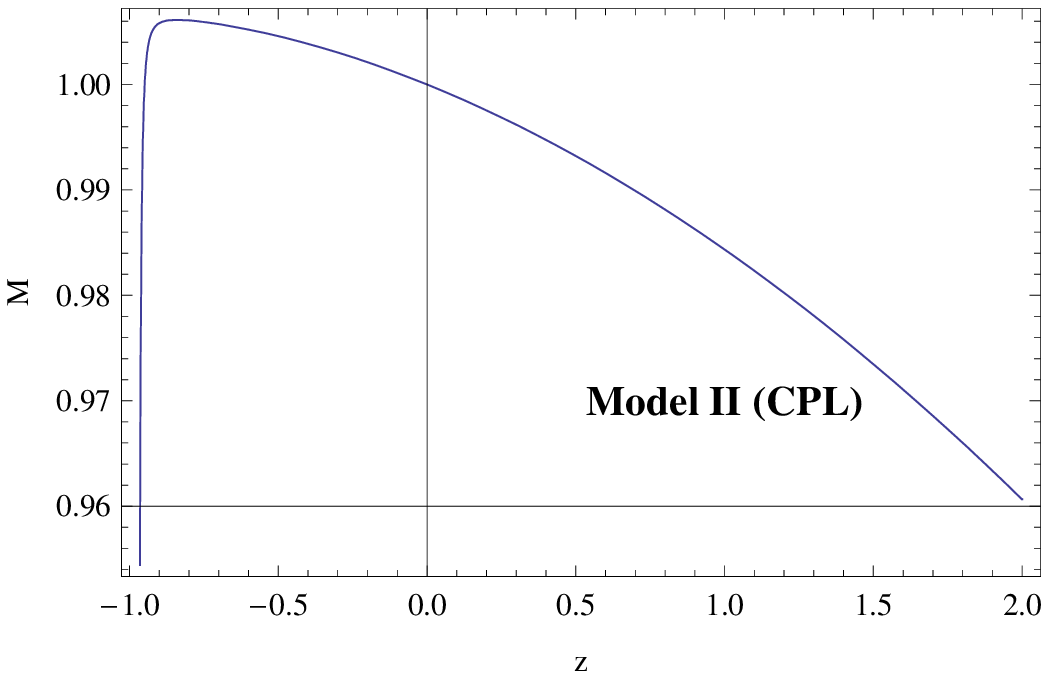}~~~~
\includegraphics[height=2.0in]{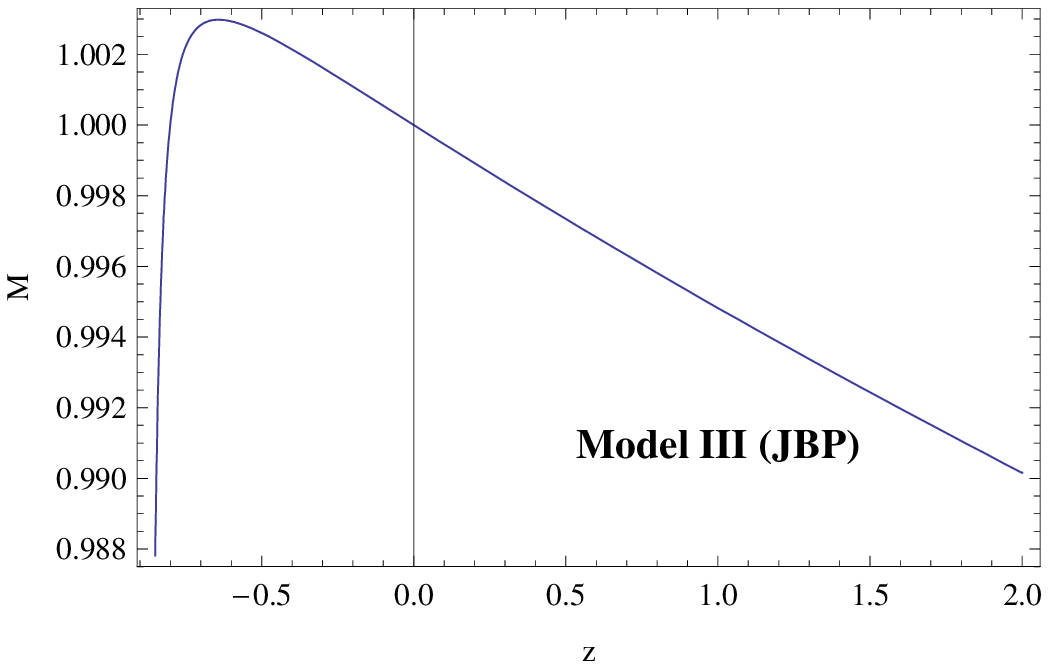}
\vspace{4mm}
~~~~~Fig.3~~~~~~~~~~~~~~~~~~~~~~~~~~~~~~~~~~~~~~~~~~~~~~~~~~~~~~~~~~~~~~~~~~~~~~Fig.4\\
\vspace{4mm}
\includegraphics[height=2.0in]{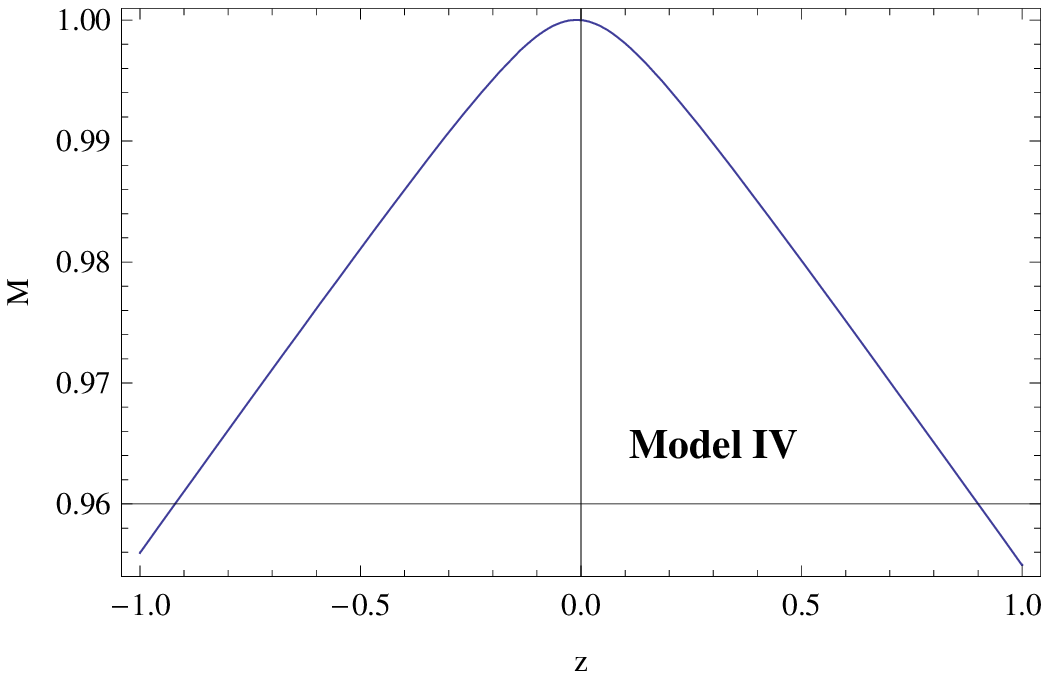}~~~~
\includegraphics[height=2.0in]{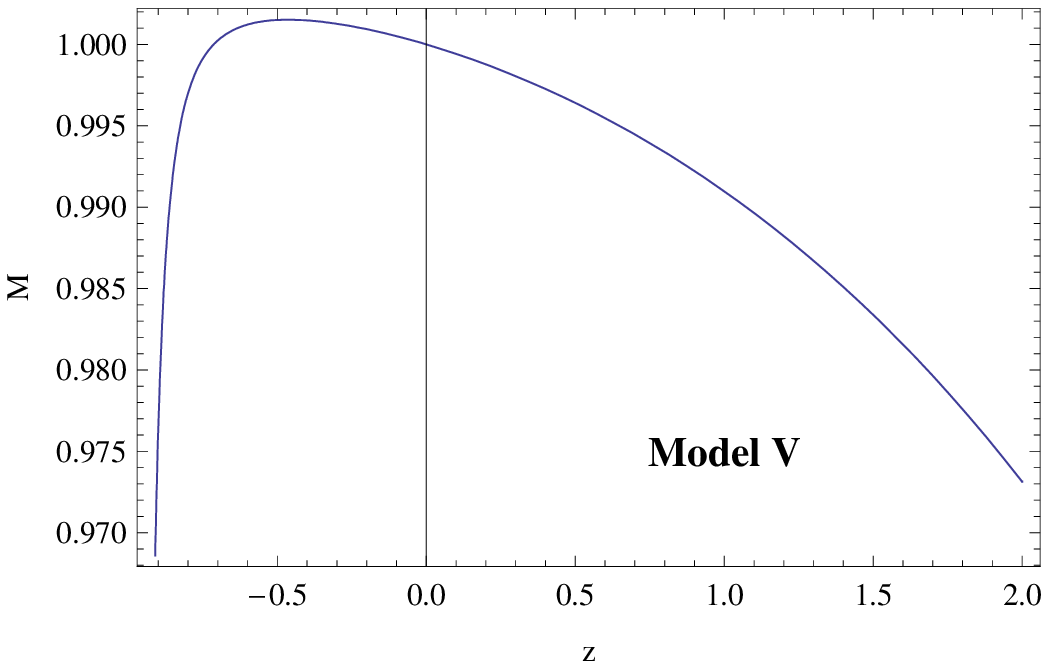}
\vspace{4mm}
~~~~~Fig.5~~~~~~~~~~~~~~~~~~~~~~~~~~~~~~~~~~~~~~~~~~~~~~~~~~~~~~~~~~~~~~~~~~~~~~Fig.6\\
\vspace{4mm}

Fig. 1 shows the variation of BTZ BH mass $M$ against redshift $z$
for MCG. Figs. 2-6 show the variations of BTZ BH mass $M$ against
redshift $z$ for Models I-V respectively. \vspace{0.2in}
\end{figure}

\subsection{Some Parameterizations of dark energy Models}

In astrophysical sense, the dark energy is popular to have a
redshift parametrization (i.e., taking the redshift $z$ as the
variable parameter of the EoS only) of the EoS as $p(z)=w(z)
\rho(z)$. The EoS parameter $w$ is currently constrained by the
distance measurements of the type Ia supernova observation with
the range of EoS as $-1.38<w<-0.82$ \cite{Melch} and WMAP3
observation to constraint on the EOS $w=-0.97^{+0.07}_{-0.09}$ for
the DE, in a flat universe \cite{Sel}. We consider following three
models of well known parametrizations (Models I, II, III). We
shall also assumed other two parametrizations (Models IV, V).
Since the following models generate both quintessence ($w(z)>-1$)
and phantom ($w(z)<-1$)
dark energies for some suitable choices of the parameters. \\

$\bullet$ {\bf Model I (Linear):} The $``Linear"$ parametrization
is given by the EoS $w(z)=w_{0}+w_{1}z$ \cite{Coor}. For Linear
parametrization and using equation (33), we get the solution as
\begin{equation}
\rho=\rho_{0}(1+z)^{3(1+w_{0}-w_{1})}e^{3w_{1}z}
\end{equation}
where, $\rho_{0}$ is the present value of the energy density. The
above model generates phantom energy if $w(z)<-1$ i.e,
$z<-\frac{1+w_{0}}{w_{1}}$ provided $w_{1}>0$ and $w_{1}-w_{0}>1$.
Using equation (2), the mass of the black hole is obtained as
\begin{eqnarray}
M=M_{0}-\frac{4\pi^{\frac{n+1}{2}}A\sqrt{\rho_{0}}}{\sqrt{3}~\Gamma(\frac{n+1}{2})}
~\left[(1+z)^{\frac{3}{2}(1+w_{0}-w_{1})}e^{\frac{3}{2}w_{1}z} -1
\right]
\end{eqnarray}

Since this model is the phantom crossing model, so if this dark
energy accretes onto BTZ BH, for quintessence era, BH mass
increases upto a certain limit and after that for phantom era, the
mass of the BH decreases. We have shown this scenario in figure 2.
From the figure, we see that BTZ BH mass $M$ increases for
redshift $z$ decreases upto certain stage of $z$ ($\Lambda$CDM
stage) and then $M$ decreases (phantom era) as universe expands.\\

$\bullet$ {\bf Model II (CPL):} $``CPL"$ parametrization
\cite{Chev,Linder} is given by the EoS
$w(z)=w_{0}+w_{1}\frac{z}{1+z}$. In this case, the solution
becomes
\begin{equation}
\rho=\rho_{0}(1+z)^{3(1+w_{0}+w_{1})}e^{-\frac{3w_{1}z}{1+z}}
\end{equation}
The above model generates phantom energy if $w(z)<-1$ i.e,
$z<-\frac{1+w_{0}}{1+w_{1}}$ provided $w_{1}>-1$ and
$w_{1}-w_{0}>0$. The mass of the black hole is obtained as
\begin{eqnarray}
M=M_{0}-\frac{4\pi^{\frac{n+1}{2}}A\sqrt{\rho_{0}}}{\sqrt{3}~\Gamma(\frac{n+1}{2})}
~\left[(1+z)^{\frac{3}{2}(1+w_{0}+w_{1})}e^{-\frac{w_{1}z}{2(1+z)}}
-1 \right]
\end{eqnarray}

This model is also the phantom crossing model. We have drawn $M$
vs $z$ in figure 3. From the figure, we observe that BTZ BH mass
$M$ increases for redshift $z$ decreases upto certain stage of $z$
($\Lambda$CDM stage) and then $M$ decreases (phantom era) as
universe expands.\\

$\bullet$ {\bf Model III (JBP):} The $``JBP"$ parametrization
\cite{Jassal} is given by the EoS
$w(z)=w_{0}+w_{1}\frac{z}{(1+z)^{2}}$. The solution is
\begin{equation}
\rho=\rho_{0}(1+z)^{3(1+w_{0})}e^{\frac{3w_{1}z^{2}}{2(1+z)^{2}}}
\end{equation}
The above model generates phantom energy if $w(z)<-1$ i.e,
$z<-1+\frac{\sqrt{4(1+w_{0})w_{1}+w_{1}^{2} } }{2(1+w_{0})}$
provided $w_{0}>-1$ and $w_{1}<-4(1+w_{0})$. The mass of the black
hole is obtained as
\begin{eqnarray}
M=M_{0}-\frac{4\pi^{\frac{n+1}{2}}A\sqrt{\rho_{0}}}{\sqrt{3}~\Gamma(\frac{n+1}{2})}
~\left[(1+z)^{\frac{3}{2}(1+w_{0})}e^{\frac{3w_{1}z^{2}}{4(1+z)^{2}}}
-1 \right]
\end{eqnarray}

This model is also the phantom crossing model. From figure 4, we
see that BTZ BH mass $M$ increases for redshift $z$ decreases upto
certain stage of $z$ ($\Lambda$CDM stage) and then $M$
decreases (phantom era) as universe expands.\\

$\bullet$ {\bf Model IV:} Another type of parametrization is
considered as in the form of EoS $ w(z)=-1+
\frac{A_{1}(1+z)+2A_{2}(1+z)^{2}}{3\left[A_{0}+A_{1}(1+z)+A_{2}(1+z)^{2}\right]}
$~~, where $A_{0},~A_{1}$ and $A_{2}$ are constants
\cite{Alam,Alam1}. This ansatz is exactly the cosmological
constant $w = -1$ for $A_{1} = A_{2} = 0$ and DE models with $w
=-2/3$ for $A_{0} = A_{2} = 0$ and $w = -1/3$ for $A_{0} = A_{1} =
0$. In this case, we get the solution
\begin{equation}
\rho=\frac{\rho_{0}[A_{0}+A_{1}(1+z)+A_{2}(1+z)^{2}]}{A_{0}+A_{1}+A_{2}}
\end{equation}
The above model generates phantom energy if $w(z)<-1$ i.e,
$z<-1-\frac{A_{1}}{A_{2}}$ provided $A_{0}<0$, $A_{1}>0$,
$A_{2}<0$ and $A_{0}+A_{1}+A_{2}<0$. For this condition, $\rho$ is
still positive. The mass of the black hole is obtained as
\begin{eqnarray}
M=M_{0}-\frac{4\pi^{\frac{n+1}{2}}A\sqrt{\rho_{0}}}{\sqrt{3}~\Gamma(\frac{n+1}{2})}
~\left[\frac{\{A_{0}+A_{1}(1+z)+A_{2}(1+z)^{2}\}^{\frac{1}{2}}}{(A_{0}+A_{1}+A_{2})^{\frac{1}{2}}}
-1 \right]
\end{eqnarray}

This model is also the phantom crossing model. From figure 5, we
see that BTZ BH mass $M$ increases for redshift $z$ decreases
upto certain stage of $z$ and then $M$ decreases (phantom era) as universe expands.\\

$\bullet$ {\bf Model V:} Other type of parametrization is assumed
to be $ w(z)=w_{0}+w_{1}~log(1+z) $~\cite{Ef,Sil}. The solution is
obtained as
\begin{equation}
\rho=\rho_{0}(1+z)^{3(1+w_{0})}e^{\frac{3}{2}w_{1}[log(1+z)]^{2}}
\end{equation}
The above model generates phantom energy if $w(z)<-1$ i.e,
$z<-1+e^{-\frac{w_{0}}{w_{1}}}$ provided $w_{1}>0$. The mass of
the black hole is obtained as
\begin{eqnarray}
M=M_{0}-\frac{4\pi^{\frac{n+1}{2}}A\sqrt{\rho_{0}}}{\sqrt{3}~\Gamma(\frac{n+1}{2})}
~\left[(1+z)^{\frac{3}{2}(1+w_{0})}e^{\frac{3}{4}w_{1}[log(1+z)]^{2}}
-1 \right]
\end{eqnarray}

This model is also the phantom crossing model. From figure 6, we
see that BTZ BH mass $M$ increases for redshift $z$ decreases upto
certain stage of $z$ and then $M$ again decreases (phantom era) as
universe expands.

\section{Discussions}

In this work, we have studied the accretion of $(n+2)$-dimensional
charged BTZ black hole (BH). A proper dark-energy accretion model
for black holes have been obtained by generalizing the Michel
theory \cite{Michel} to the case of black holes. Such a
generalization has been already performed by Babichev et al
\cite{Babichev,Babichev1} for the case of dark-energy accretion
onto Schwarzschild black holes. We have followed the procedure
used by Babichev et al \cite{Babichev,Babichev1}, adapting it to
the case of $(n+2)$-dimensional charged BTZ black hole. The
critical point and square speed of sound have been obtained.
Astrophysically, mass of the black hole is a dynamical quantity,
so the nature of the mass function is important in our black hole
model for different dark energy filled universe. We see that the
rate for the BTZ black hole exotic mass due to accretion of dark
energy becomes exactly the positive to the similar rate in the
case of a Schwarzschild black hole, asymptotically. Since the BTZ
black hole is static, so the mass of the black hole depends on $r$
only. When some fluid accretes outside black hole, the mass
function $M$ of the black hole is considered as a dynamical mass
function and hence it should be a function of time also. So
$\dot{M}$ is time dependent and the increasing or decreasing of
the black hole mass $M$ sensitively depends on the nature of the
fluid which accretes upon the black hole. The sign of time
derivative of black hole mass depends on the signs of $(\rho+p)$.
If $\rho+p<0$ i.e., for phantom dark energy accretion, the mass of
the black hole decreases but if $\rho+p>0$ i.e., for quintessence
dark energy accretion, the mass of the black hole increases. The
mass of the BTZ BH has been calculated and we have observed that
the mass of the BTZ BH is related with square root of the energy
density of dark energy which accretes onto BH in our accelerating
FRW universe.\\

We have assumed modified Chaplygin gas (MCG) as a candidate of
dark energy which accretes onto BTZ BH. Since in our solution of
MCG, this model generates only quintessence dark energy (not
phantom) and so BTZ BH mass increases during the whole evolution
of the accelerating universe, which is shown in figure 1 also.
Next we have assumed 5 kinds of parametrizations (Models I-V) of
well known dark energy models (some of them are Linear, CPL, JBP
models). These models generate both quintessence and phantom
scenarios (phantom crossing models) for some restrictions of the
parameters. So if these dark energies accrete onto the BTZ black
hole, then for quintessence stage, black hole mass increases upto
a certain value (finite value) and then decreases to finite value
for phantom stage during whole evolution of the universe. That
means, if the 5 kinds of DE accrete onto BTZ black hole, the mass
of the black hole increases upto a certain finite value and then
decreases in the late stage of the evolution of the universe. We
also shown these results graphically clearly. We have drawn the
mass of the BTZ black hole for dark energy models I-V in figures
2-6 respectively. Figures 2-6 show the mass of the BTZ black hole
first increases to finite value and then
decreases to a finite value also.\\

\end{document}